\documentclass[conference]{IEEEtran}
\usepackage{cite}
\usepackage{amsmath}
\usepackage{amsfonts}
\usepackage{setspace}
\usepackage{amsthm}
\usepackage{tabularx}
\usepackage{graphicx}
\usepackage{enumerate}
\usepackage{multirow}
\usepackage{empheq}
\usepackage{textcase}
\usepackage{algorithmic}
\usepackage{textcomp}
\usepackage{xcolor}
\usepackage{amsmath,amssymb,amsfonts}

\IEEEoverridecommandlockouts

\DeclareRobustCommand{\IEEEauthorrefmark}[1]{\smash{\textsuperscript{\footnotesize #1}}}

\begin{document}

\title{Drive Right: Shaping Public's Trust, Understanding, and Preference Towards Autonomous Vehicles Using a Virtual Reality Driving Simulator

\thanks{This work received support from Mobility 21 University Transportation Center at Carnegie Mellon University (Project 342) which was sponsored by the U.S. Department of Transportation.}
}

\makeatletter 
\newcommand{\linebreakand}{%
  \end{@IEEEauthorhalign}
  \hfill\mbox{}\par
  \mbox{}\hfill\begin{@IEEEauthorhalign}
}
\makeatother 

\author{
    \IEEEauthorblockN{Zhijie Qiao\IEEEauthorrefmark{1},
    Xiatao Sun\IEEEauthorrefmark{1},
    Helen Loeb\IEEEauthorrefmark{2},
    Rahul Mangharam\IEEEauthorrefmark{1}
    }
    \IEEEauthorblockA{
        \IEEEauthorrefmark{1}Department of Electrical and Systems Engineering\\
        University of Pennsylvania, Philadelphia, PA, 19104\\
        Email: zhijie@alumni.upenn.edu
    }
    \IEEEauthorblockA{
        \IEEEauthorrefmark{2}Jitsik LLC, 276 Barwynne Road, Wynnewood, PA, 19096
    }
}

\maketitle


\begin{abstract}
Autonomous vehicles are increasingly introduced into our lives. Yet, people’s misunderstanding and mistrust have become the major obstacles to the use of these technologies. In response to this problem, proper work must be done to increase public’s understanding and awareness and help drivers rationally evaluate the system. The method proposed in this paper is a virtual reality driving simulator which serves as a low-cost platform for autonomous vehicle demonstration and education. To test the validity of the platform, we recruited 36 participants and conducted a test training drive using three different scenarios. The results show that our simulator successfully increased participants’ understanding while favorably changing their attitude towards the autonomous system. The methodology and findings presented in this paper can be further explored by driving schools, auto manufacturers, and policy makers, to improve training for autonomous vehicles.
\end{abstract}



\section{INTRODUCTION}

\subsection{Background \& Motivation}

Autonomous vehicles (AVs) have been gaining unprecedented attention as the society eagerly expects a revolution in the transportation industry. While this technology will require a true leap of faith from drivers, little work has been done to increase the public’s understanding and awareness. Research has shown that the public has a natural tendency to resist AVs as it expects them to be much safer than human driven vehicles before considering their use \cite{LIU2020692, Muhl20201322}. Similarly, AV related malfunctions and accidents tend to be overly dramatized in the media which further jeopardizes public opinion \cite{LIU2019232}.

As we continue with the technological development of AVs, it is unlikely that in the near future, AVs will be universally adopted and completely outperform human drivers. Yet, while the SAE level 5 of automation still need years of development, drivers could already take advantage of the assistance features at level 1 and 2. Models based on economic and technological prediction have shown that by 2045, level 1 autonomy will be adopted by over 90 percent of the U.S. vehicle fleets but the adoption rate of level 4 varies from 24.8 percent to 87.2 percent \cite{BANSAL201749}. Despite the social, economic, and technological challenges, AVs have the potential to: significantly reduce traffic accidents, increase traffic efficiency, add environmental benefits, and provide increased mobility options \cite{HASAN2021102929, 10.1007/978-3-319-93885-1_29}.

In order for society to fully accept AVs and enjoy their benefits, work must be done to increase the public’s understanding and trust \cite{Abraham2017}. However, ``it is not enough to just trust and use AVs ... one must trust them \emph{appropriately} and use them \emph{ properly}" \cite{Wintersberger2019}. Blinded mistrust towards AVs will cause a waste of their benefits but undoubted trust can lead to serious and even fatal accidents \cite{KAYE2021352}. Trösterer et al points out that lessons can be learned from the pilots' interaction with the flying automation system. From their study, one pilot claimed that ``The automatic functions are very, very reliable, very, very good, very, very sensitive as well, but not 100\%" \cite{10.1145/3122986.3123020}. Therefore, it is reasonable to suggest that, in the current condition of imperfect AVs, a certain amount of alertness and training will be needed. To that extent, Revell et al. explored the Perceptual Cycle Model by having six UK drivers interact with a Mercedes S Class equipped with level 2 assistance features in the real traffic. A lot of drivers’ frustration was observed and there was clearly a lack of synergy between the drivers and the vehicle \cite{REVELL2020103037}. This further shows that training can play a beneficial role during drivers' initial interaction with AVs. However, one must remember that on-road training carries some level of risk, and potentially requires a safety operator, which can be costly and time-consuming. Moreover, such training is subject to traffic, weather, and geographical conditions and may not provide the most comprehensive evaluation \cite{KALRA2016182}.

\subsection{Simulator Study}

To solve the above problems, we propose a driving simulator which stands up as a cost-effective, time-efficient, and completely safe platform to help drivers get familiarized with the capabilities of AVs. Driving simulators allow for the testing of different scenarios that might be too dangerous or risky to attempt in the real world \cite{MALIK2022742}. They can be combined with deep learning and machine learning models to create probabilistic rare case events with high accuracy \cite{Wen2020, 10.5555/3327546.3327650}. They are also accessible to people from all backgrounds and with varying driving skills.

This work is a continuation of our previous drive right effort which aimed at increasing public’s understanding and trust towards AVs using a simulation educational platform \cite{12-05-04-0028}. Our previous study has shown that the simulator education could effectively decrease participants’ perceived risk and increase the perceived usefulness of AVs. In this study, we developed a virtual reality (VR) driving simulator based on the open-source simulator CARLA \cite{Dosovitskiy2017} which allows the user to sit in the simulator chair, wearing an Oculus Quest 2 headset, and interact with the vehicle as if it was a real one (Fig. \ref{fig:demo}). A pilot human study was designed and a corresponding survey was developed in an effort to investigate two research questions: 1. Does a driving simulator systematically change the drivers’ attitude to be more positive towards AVs and increase their intention to use one? 2. Can a driving simulator be a good tool for AV education and demonstration? To that effect we recruited 36 licensed drivers for an interactive experiment and summarized the findings in the next few sections.

\begin{figure}[t]
     \centering
     \includegraphics[width=0.85\linewidth]{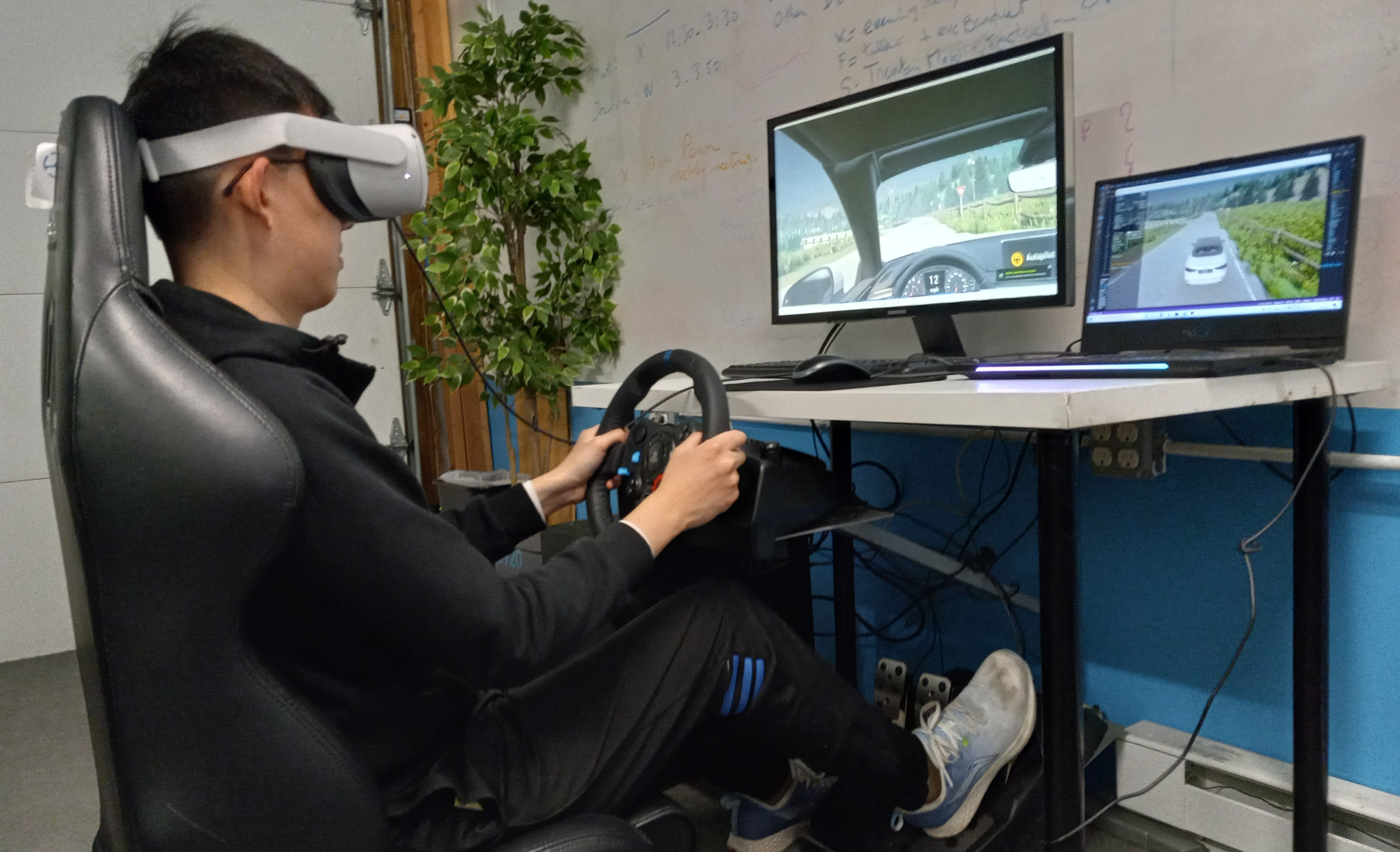}
     \caption{Simulator demonstration: left monitor shows CARLA server first person driver’s view which is also displayed in the VR headset; right monitor shows the CARLA client third-person tracking view which is provided by the CARLA client by default.}
     \label{fig:demo}
\end{figure}

\subsection{Contribution}
To the best of our knowledge, this work makes the following contributes to the existing field:

\begin{itemize}
    \item It is the first rigorous approach to deploy the CARLA driving simulator into the VR framework. This effort allows us to extend its driving algorithm validation purpose with a human factor consideration. Our development provides an easy-to-follow approach and facilitates the use of CARLA on any VR headset. Full instructions are available on our lab website \cite{mlab2022}.
    \item It is the first attempt with a holistic approach to use AV systems that focus entirely on user experience and interaction. In our simulator, everything is designed from a user standpoint to help drivers form their own understanding of AVs in a vivid and immersive environment. 
    \item We present the rationale of using a driving simulator 
    and show why it can be such an effective tool for AV training and demonstration. It is our hope that the results from our study will provide some guidance to auto dealerships, sales forces, driving schools instructors, and policy makers.
\end{itemize}

\section{Literature Review}

A great amount of effort has been made to influence the public’s attitude towards AVs using a simulator. The goal of our review is to explore what level of information and interactivity can inspire user adoption and maximize the benefit of using an AV. In a general sense, these experiments can be categorized into two groups: passive demonstration and active interaction.

In a passive demonstration, the user receives information from the autonomous system through various forms, but does not attempt to take control of the vehicle. In \cite{9353546, MA2021103272, 10.1145/3411764.3446647}, researchers displayed pre-recorded real traffic scenarios with animation and visual enhancement to mimic the condition of an AV. The surrounding environment including cars, pedestrians, traffic signs and the AV’s planned movement were marked up for explanation with a preview of the AV's actions. For example, Ha et al. manually controlled and recorded the vehicle in a city car simulator. They showed the VR environment to the participants while adding different levels of explanation. The participants could change their focus and point of view using the VR headset but could not interfere with the AV control \cite{Ha2020271}. This approach has been shown to be effective as the user can often gain a good understanding of the AV’s working mechanism and form a rational expectation of its actions. However, one drawback is that the excessive information can cause increased user stress and mental workload, which partially offset the benefits of using an AV. 

Another popular choice for information delivery is anthropomorphism, where the AV interface is designed with human features, thus portraying the system as intelligent and trustworthy \cite{10.1007/978-3-642-39182-8_9}. In \cite{Niu2018352}, the AV’s steering, acceleration, and state information were represented using the robots’ eye color, movement, and blinking rate. In \cite{LEE2022107015}, the AV used polite speech strategy to gain drivers' trust and favor, while in \cite{mti2040062}, a combination of graphical and conversational user interface was used. Results have shown that the anthropomorphic interfaces can be an effective way to increase drivers' confidence and trust in AVs, and consequently, facilitate the adoption of the technology. Nevertheless, The best way to implement the anthropomorphic design is still up for debate. 

In an active simulator setup, participants not only receive information but also actively try to interact with and take control of the AV. The technology presented is usually an imperfect or semi-AV, but the research scope varies. For instance, Park et al. explored drivers' adaptability and preference towards full and semi-AVs using a fixed based simulator. They found that quick interaction experience reduces anxiety and potential loss of control \cite{8705036}. Shi et al. collected data on the AV’s lateral position, steering wheel angle mean, and standard deviation to figure out the effect of driving styles on semi-AVs \cite{shi2019}. Ebnali et al. found that simulator training can significantly increase users' performance on semi-AVs in several aspects such as the takeover time, speed, decision accuracy, trust, and acceptance. A later experiment across three different platforms showed that high-fidelity VR is the most effective tool for AV demonstration and familiarization \cite{EBNALI2019184, EBNALI2021103226}. 

\begin{figure}[t]
     \centering
     \includegraphics[width=0.9\linewidth]{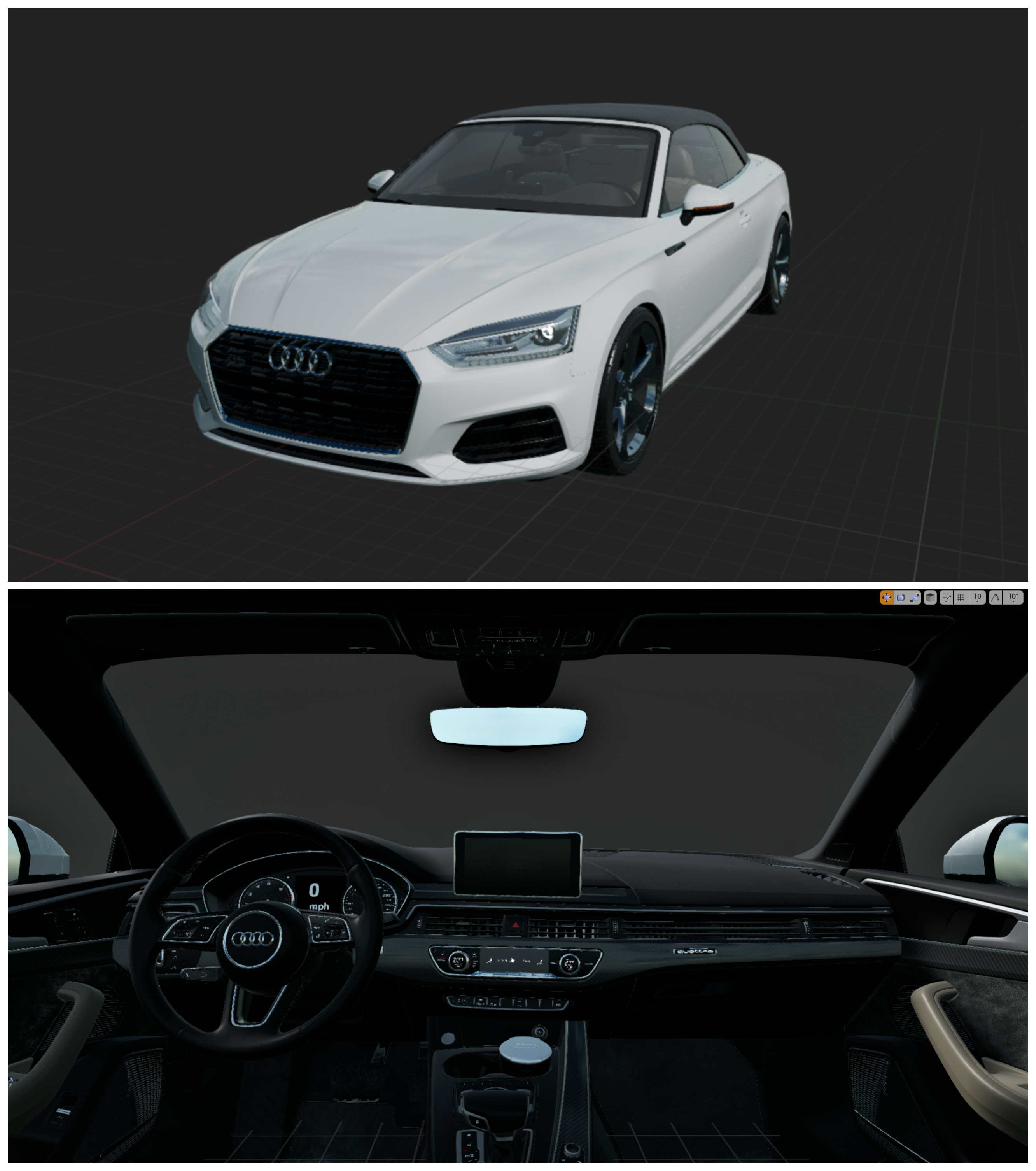}
     \caption{High level of detail Audi vehicle 3D model exterior and interior view.}
     \label{fig:carview}
\end{figure}

\section{Design}

\subsection{CARLA}

Various AV simulators have been developed by academia and industry, of which, two state-of-the art ones support an open license: CARLA \cite{Dosovitskiy2017} and the SVL Simulator \cite{rong2020lgsvl}. The SVL Simulator suspended its service in January, 2022, while the CARLA platform provides continuous support and updates. CARLA was developed from the ground up to support the training and validation of an autonomous driving system. It has a rigorous collection of digital assets (maps, vehicles, pedestrians) and a complete sensor suite (Camera, RADAR, LiDAR, GNSS, IMU). With a flexible API and the integration of ROS bridge, CARLA can be used to test various subsystems of an AV \cite{MALIK2022742}. In this study, we extended CARLA's driving algorithm validation focus with a human factor consideration. By taking advantage of CARLA's well constructed environment and the immersiveness of a VR headset, we reduce the gap between the simulator and a real vehicle to support human-centered AV studies.

\subsection{CARLA VR}

While the standalone CARLA package is built in the 3D environment, it does not come with the VR capability. Therefore, the research team added the OpenXR plugin and rebuilt the project to enable rendering on an Oculus Quest 2 headset. To achieve this, the CARLA project source code and a CARLA-customized fork of Unreal Engine 4 (UE4) were downloaded from the official platform. Then, the UE4 was built using the Visual Studio 2019, Windows 8.1 SDK, x64 Visual C++ Toolset, .NET framework 4.6.2. The CARLA server and client were compiled using the x64 Native Command Tool. In this setup, the project can easily accommodate another headset in the future by replacing the path and performing the re-compilation.

The default CARLA simulator only shows the vehicles in the third-person’s view with a low level of detail for mesh and texture. To use CARLA in the VR application in the first-person view, a high level of detail vehicle with interior modeling must be added. After searching and comparing a variety of models, we selected an Audi A6 model from the Car Configurator. This asset was developed by Epic Games and supports a free license (Fig. \ref{fig:carview}) when working with UE4. It decouples rendering and collision detection by using a high-fidelity vehicle mesh while leveraging the simplified polygons for collision check, which ensures high rendering quality and significantly saves the computing power, allowing the simulator to run smoothly. 

A Logitech G29 steering wheel and pedal set is used alongside the simulator. It supports two different modes: manual and autonomous. In the manual mode, the user uses the steering wheel to control the direction of the vehicle, and the throttle and brake pedals to adjust the speed. The vehicle is assumed to have automatic gear. Systematic tests showed that the elapsed time from the user input to the simulator response took less than 100 microseconds, which is below the human reaction time. The autonomous mode can be triggered by pulling the lever on the left side of the steering wheel. To exit, the user can toggle the lever again or lightly press the brake pedal. Note that in both the manual and autonomous mode, the physical steering wheel perfectly mimics the position of the virtual wheel in the simulated car model. This helps reduce any sensing discrepancy between the physical wheel and the simulation while minimizing the discontinuity during a driving mode switch (Fig. \ref{fig:steer_wheel}). 

\subsection{Driving Mode}

The goal of the study is to let drivers experience AVs in an environment that feels as real as possible. The focus is not to help designers improve the system but to help drivers understand what it feels like to ride in an AV, how they can interact with it, so that they can decide whether an AV is a good fit for them. To that purpose, we developed two driving modes in our simulator: manual and level 4 autonomous. By defining that the system is level 4, the AV could: take in a destination input and navigate to that place without intervention. While traveling, the AV is capable of: adaptively adjusting its speed, performing emergency yield or brake, and taking over other vehicles while maintaining safe distances. The autonomous algorithms were developed based on CARLA's built-in modules but modified to support more intelligent and reasonable behavior. 

During the drive, the driver could switch between the manual and autonomous mode at any time. The researcher did not attempt to gain complete control of the experiment, but instead let each driver decide if they wanted to use the autonomous mode and how they would like to use it. If an autonomous mode was triggered in the case of an imminent emergency, the system would try to make its best decision to either yield or brake and steer the vehicle back to safety. Note that our AV was set at level 4 mainly to fulfill the goal of education and demonstration. If the system was set at level 5, then there is no need for human interaction. On the contrary, if the system was designed for level 2 or 3, it would require constant human intervention, which jeopardizes user's trust and violates the initiative of the study. For future development where a more faithful AV stack is required, CARLA can be linked with the open-source software such as the Autoware \cite{Autoware2022} and Apollo \cite{apollo2022}. This will allow the vehicle to perform in the simulator as it would do in real life.

\begin{figure}[t]
     \centering
     \includegraphics[width=0.9\linewidth, ]{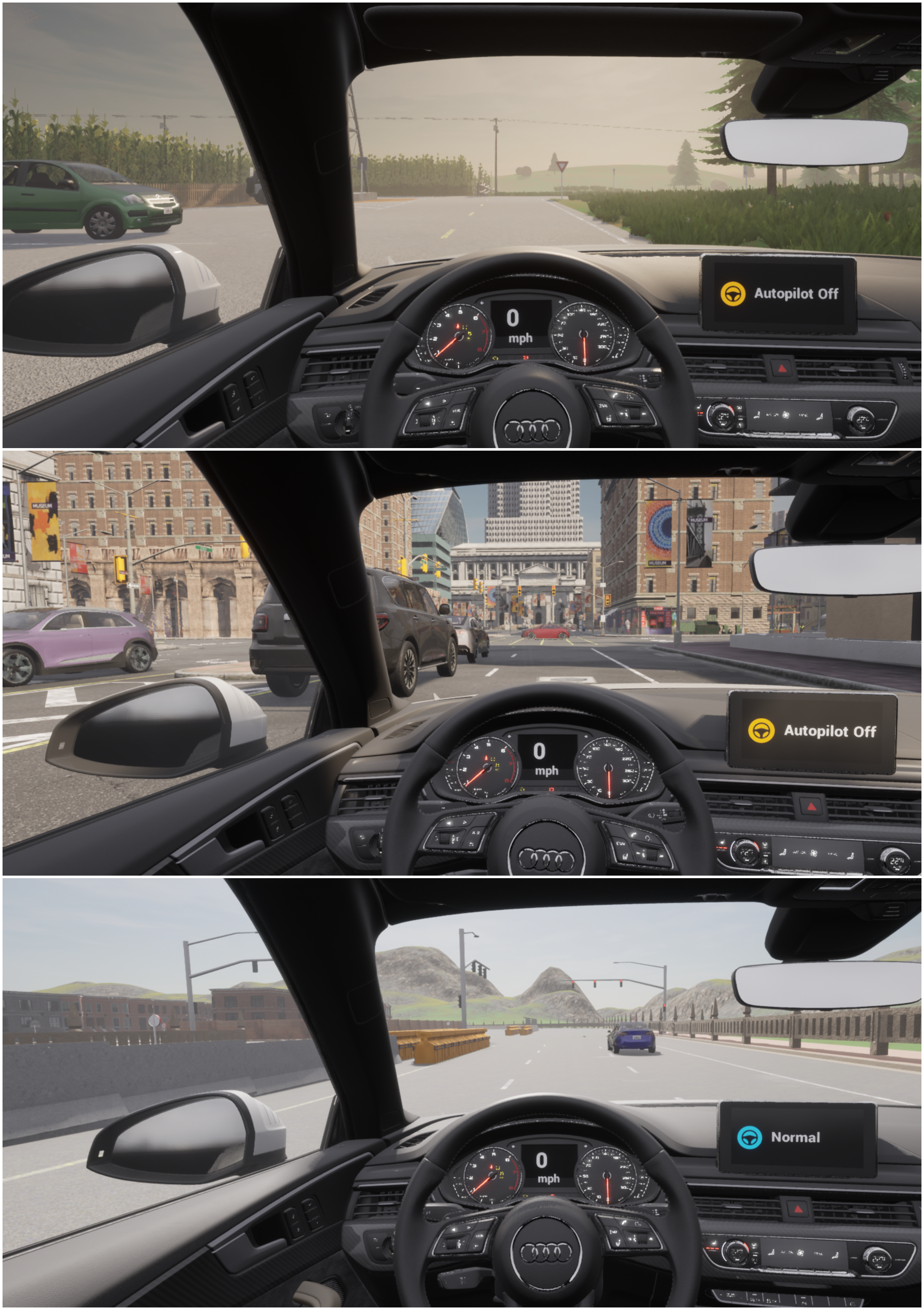}
     \caption{Rural, city, and highway scenarios in the first person driver's view (from top to bottom).}
     \label{fig:scenarios}
\end{figure}

\subsection{Scenarios}

Three scenarios were designed in our simulator: rural, city, and highway. All scenario maps were imported from the CARLA environmental asset with no further modification. In addition, they were designed so as to guide users to gradually get used to the interaction with the AV, understand its capabilities, and start building confidence in it (Fig. \ref{fig:scenarios}).

The rural scenario is a representation of the suburban environment, and consists mostly of trees, shrubs, one-lane country roads, and stop signs at intersections. This scenario serves as a familiarization step, in which the participants actively control the vehicle with the steering wheel and pedals in the manual mode, and let the vehicle drive itself in the autonomous mode. Since the rural scenario enforces simple road conditions and low vehicle speed, participants can get used to the simulation environment and vehicle controls in a relaxed setting. No other vehicles are added into the map to help keep the tutorial clean and simple. 

The city scenario is populated with buildings, traffic lights, and complex road crossings. In this scenario, participants are not required to drive, and are asked to turn on the autonomous mode and watch the vehicle navigate by itself. The predefined route guides the AV as it navigates several conditions such as sharing the road with conventional vehicles, merging at an unprotected intersection, waiting for traffic signals, and yielding to pedestrians. Participants can resume manual control at any time if they choose to handle an emergency situation. The goal of this scenario is to show the capabilities of an AV in a complex city environment to help participants further increase understanding and trust. 

The highway scenario is a three lane express road with no crossings or cyclists. Several NPC vehicles were added with varying velocities and driving behaviors. In this scenario, participants can switch freely between manual and autonomous mode. In addition, the AV supports three different behaviors: cautious, normal, and aggressive. Different driving behaviors have different acceleration, brake, and velocity profiles. In all cases, the AV can automatically take over another vehicle that is driving at a lower speed. For this scenario, participants were encouraged to use AV to experience its different behaviors, and they could change the behavior using the up \& down buttons on the steering wheel. These behaviors were not available in the previous scenarios primarily for two reasons. First, in the rural and city scenario, the vehicle has a relatively low speed limit and different driving styles will not show much difference. Second, we want to increase the complexity of control one step at a time so that users would not get overwhelmed by excessive simulation features.

\section{EXPERIMENTS}

\begin{table*}[t]
    \centering
    \caption{Fifteen Quantitative Survey Questions.}
    \label{tab:survey}
    \begin{tabular}{|m{2.0em}|m{46em}|}
        \hline
        TR1 & I believe that autonomous vehicles can take me safely to my destination.\\
        \hline
        TR2 & I believe that autonomous vehicles can handle most traffic conditions.\\
        \hline
        TR3 & I believe that autonomous vehicles are as reliable as my own driving.\\
        \hline
        PR1 & I am worried about the safety of autonomous vehicle technology.\\
        \hline
        PR2 & I am worried about the interaction of an autonomous vehicle with conventional vehicles.\\
        \hline
        PR3 & I am worried that autonomous system failure or malfunction may cause accidents.\\
        \hline
        PU1 & Using an autonomous vehicle will allow me to conduct non-driving related tasks.\\
        \hline
        PU2 & Using an autonomous vehicle will increase my driving safety and efficiency.\\
        \hline
        PU3 & Using an autonomous vehicle will be useful when I am physically or mentally impaired.\\
        \hline
        PE1 & Learning to operate an autonomous vehicle would be easy for me.\\
        \hline
        PE2 & Interacting with an autonomous vehicle would not require a lot of my mental effort.\\
        \hline
        PE3 & I think it is easy to get an autonomous vehicle to do what I want to do.\\
        \hline
        BI1 & I intend to ride in an autonomous vehicle in the future.\\
        \hline
        BI2 & I expect to purchase an autonomous vehicle in the future.\\
        \hline
        BI3 & I plan to introduce autonomous vehicles to my family and friends.\\
        \hline
    \end{tabular}
\end{table*}

\subsection{Participants}

36 participants were recruited via email and social media. In order to qualify for the study, they had to meet the following requirements: be older than 18 years old and less than 75; have a valid driver’s license and at least three months of independent driving experience; have no police-reported crash within the last year; have normal or correct-to-normal vision and hearing (contact lens allowed); have no history of migraine, claustrophobia, or motion sickness; not be pregnant; and have no prior interaction with AVs level 3 or above.

Our participant group consisted of 18 males, 18 females and had a mean age of 25.5 years. Participants were recruited via convenience sampling from the university community and were mostly students. A few elderly people were part of the study. The participants were compensated with a \$25 Amazon gift card for taking part in the study. If they decided to quit in the middle of the study for any reason, they were still compensated. This study was approved by the Institutional Review Board at the University of Pennsylvania (IRB Protocol\#: 850824) and all participants gave their informed written consent.

\subsection{Survey} \label{Survey}
The effectiveness of our simulator was measured through a survey instrument with quantitative and qualitative questions. The quantitative part was based on several models, including the Technology Acceptance Model (TAM) \cite{Davis1989319}, Autonomous Vehicle Acceptance Model (AVAM) \cite{10.1145/3301275.3302268}, and the Universal Theory of Usage and Acceptance of Technology (UTAUT) \cite{Venkatesh2003425}. Questions were developed with regard to five categories: Trust (TR), Perceived Risk (PR), Perceived Usefulness (PU), Perceived Ease-of-Use (PE), and Behavioral Intention (BI). Each category consisted of three questions which were evaluated using the five-point Likert scale with 1 being \emph{strongly disagree} and 5 being \emph{strongly agree}, while the category is not explicitly shown (Table \ref{tab:survey}).

TR has been identified as a key element that influences human-machine interaction, with a direct link to people’s mental adoption of the system and their intention to use it. Different studies have pointed out that a lack of TR is the most direct reason for not accepting the AV technology. In that sense, a favorable initial TR is critical in overcoming the potential risks and promoting the adoption of AVs \cite{ZHANG2019207}. PR refers to the uncertainty in a given situation, and in an AV case, unexpected behavior and loss of control. PR has been shown negatively related to TR, and a strong PR of a negative situation discourages the use of the technology \cite{doi:10.1080/10447318.2015.1070549}. PU can be explained as the positive use-performance, while PE represents the level of effort required to use a specific system. Both measures can shape an individual’s favorable opinion towards a technology \cite{NASTJUK2020120319}. BI is an individual’s intention to use an AV and it has been shown that TR, PU, and PE have a positive association with BI while PR has a negative relationship \cite{DIRSEHAN2020101361, 10.1145/2667317.2667330}.

The qualitative part focused on the participants’ subjective feedback, including their experience using our driving simulator, their opinion of using a driving simulator for auto dealership demonstrations, and their view on using a simulator at driving schools to assist with the conventional vehicle education. Participants’ were encouraged to raise any additional questions or share other comments in this section.

\subsection{Procedure}
Participants who contacted us and met the requirements were invited to our research lab. Upon arrival, they read and signed the consent form, and filled in the first part of the survey including the fifteen quantitative questions from five categories. Then, they watched a three-minute video introduction to the five levels of automation which was an explanation of the vehicle’s capabilities \cite{GeospatialWorld2018}. The video was carefully selected so that it did not affect the participants' attitude towards AV neither positively nor negatively. With this background information, the researcher then explained that the simulator AV was set at level 4. Participants then took a seat in the simulator setup as shown in Fig. \ref{fig:demo} and tried all three scenarios. In each scenario, they were asked to drive along a predefined route, but with the freedom to switch between the manual and autonomous mode at any time. After the participants finished all three scenarios, we administered the second part of the survey, which carried the same fifteen quantitative questions plus the qualitative part.

\section{Analysis}

\subsection{Wilcoxon Signed-Rank Test}
The means and standard deviations of the five categorical measurements before and after the simulator were shown in Table \ref{tab:result}. For each category, the value is an unweighted average of its three questions. In addition, Cronbach’s $\alpha$ coefficient was computed to validate the internal consistency of each category. The results have shown that the average ratings for all categories improved after the simulator demonstration, which means that participants’ opinions towards AVs in general shifted towards the positive direction.

To verify that the changes in participants’ attitude were statistically significant, a Paired-Samples T Test was initially proposed. However, a Chi-Square Test showed strong evidence $(p<0.01)$ that the difference between the experimental data was not normally distributed and violated the test assumption. Therefore, a Wilcoxon Signed-Rank Test was applied to test the data statistical significance. It is shown that after the simulator experiment, there was a significant increase in TR $(N=27,p<0.01)$, a significant decrease in PR $(N=33,p<0.05)$, a significant increase in PU $(N=31,p<0.05)$, a significant increase in PE $(N=28,p<0.01)$, and a significant increase in BI $(N=21,p<0.01)$. The significant improvement in all categories confirmed our hypothesis on the effectiveness of the simulator. 

\begin{figure}[t]
     \centering
     \includegraphics[width=0.8\linewidth]{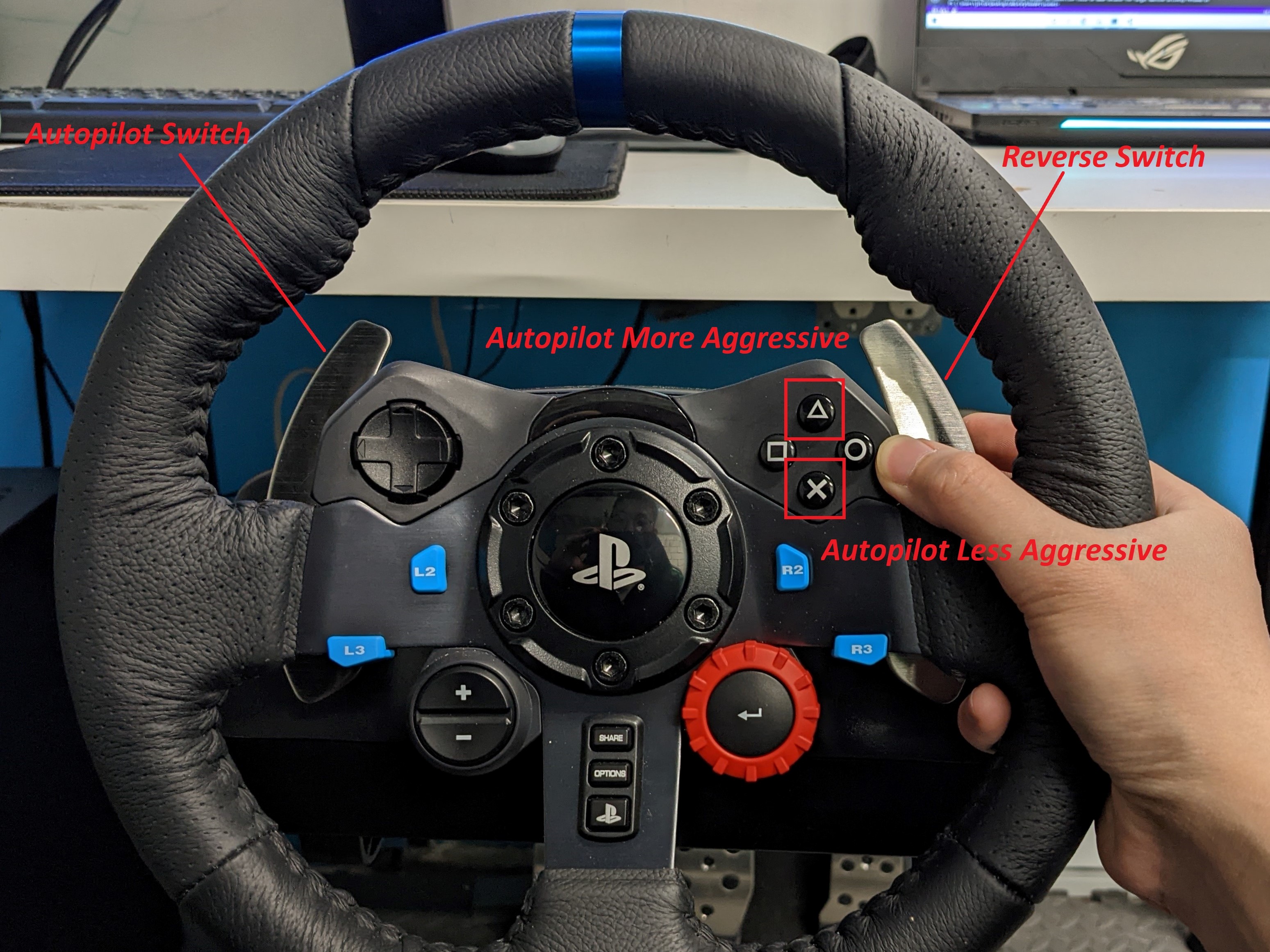}
     \caption{Logitech steering wheel functionality explanation.}
     \label{fig:steer_wheel}
\end{figure}

\subsection{Internal Consistency}
The Cronbach’s $\alpha$ coefficient showed different results before and after the simulator experiment but remained mostly consistent for each category. TR received a score of 0.75 both before and after the simulator study, meaning that trust is a consistent measure and can be bonded to a certain standard. PR initially received a score of 0.69 and increased to 0.82, indicating that participants’ evaluations on risk became better aligned across different perspectives. PE received a score of 0.52 before and 0.57 after, showing that the internal connection among the questions is only moderate and does not change much due to the simulator. Similarly, BI received high scores around 0.8, suggesting that participants’ opinions were aligned in terms of purchasing, riding, or introducing an AV. The counter-intuitive part lied in PU, which initially showed no connections among the questions with a score of 0.03 but increased to 0.67 after the simulator trial. To explain this, we took a deeper look and found that many participants originally gave a score of one for PU1, which asked about AV assisting them conducting non-driving related tasks. In comparison, they gave high scores for PU2: AV helps increase driving safety and efficiency, and PU3: AV helps drivers that are physically and mentally impaired. The high inconsistency showed that although participants agreed on AV’s safety assistance, they were concerned in handing over control while focusing on other tasks. After the simulator experiment, however, participants built a sense of what AV felt like and became more confident in its performance, which helped align the measures and contributed to a higher Cronbach’s $\alpha$. 

\begin{table}[t]
    \centering
    \caption{Mean, Standard Deviation, and Cronbach’s $\alpha$ of the Five Categorical Measurements}
    \label{tab:result}
    \begin{tabular}{            |m{1.5em}|m{2em}|m{2em}|m{2em}|m{2em}|m{2em}|m{2em}|}
        \hline
        & \multicolumn{3}{|c|}{Pre Simulator} & \multicolumn{3}{|c|}{Post Simulator}\\
        \hline
        & M & SD & $\alpha$ & M & SD & $\alpha$\\
        \hline
        TR & 3.12 & 0.77 & 0.75 & 3.60 & 0.78 & 0.75\\ 
        \hline
        PR & 3.44 & 0.75 & 0.69 & 2.95 & 0.84 & 0.82\\
        \hline
        PU & 3.53 & 0.65 & 0.03 & 3.90 & 0.79 & 0.67\\
        \hline
        PE & 3.35 & 0.73 & 0.57 & 3.72 & 0.71 & 0.52\\ 
        \hline
        BI & 3.63 & 0.84 & 0.78 & 3.98 & 0.76 & 0.82\\ 
        \hline
    \end{tabular}
\end{table}

\subsection{Correlation}
Our results show that TR has a moderate positive correlation with BI (0.457 before the simulator and 0.535 after the simulator), PR has a moderate negative correlation with BI (-0.385 before and -0.236 after), PU has a moderate positive correlation with BI (0.381 before and 0.315 after), and PE has a moderate positive correlation with BI (0.355 before and 0.317 after). None of the correlation coefficients appear to be quite weak or strong, but all seem consistent before and after the simulator study with some reasonable fluctuation. The result matches the literature review and survey design as shown in Section \ref{Survey}.

\section{Discussion}

\subsection{AV Behavior Choice}
After the highway mode, participants were asked about whether they liked to stick to one AV behavior or use all three behaviors (cautious, normal, aggressive) interchangeably. About half of the participants stated that they prefer the availability of all three behaviors, so that they could decide which one to use based on the timing of things, safety perception , and level of comfort. Another group answered that it would be nice to have all three choices, but they would almost always use the aggressive one. The remaining participants said that they just wanted aggressive behavior, and according to them, if the AV was designed to prioritize safety, “why not drive as fast as you can?” 

\subsection{Driver’s Background}
We also found that drivers’ evaluations on AVs were dependent on their own perceived driving skills. Participants who claimed to have a lot of driving experience tended to not trust the AV: they described the AV as safe but rigid, and definitely not a match for their own driving. On the contrary, participants who drove little and were concerned with their own driving safety showed higher interest in AV and praised its safety feature. From this observation, we can reasonably infer that people who are more dependent on AV and who can benefit most from it will be more willing to embrace the technology, while people who heavily rely on and enjoy their own driving will be reluctant to accept it. This finding is also supported by \cite{Garidis2020} which states that ``\emph{loss of driving pleasure}" and ``\emph{the desire to exert control}" make drivers more negative towards AVs \cite{HERRENKIND2019255}.

\subsection{Simulator for AV Demonstration}
The researchers discussed with the participants the use of the simulator for AV demonstration. A typical example would be at auto dealerships, and most participants stated that they definitely would want some simulation experience to help them make the choice. While it is agreed that driving simulators cannot provide the physical motion and the sense of alertness as in real traffic, they come with their unique advantages. First, a driving simulator is not constrained by the available conditions during a test drive, and grants the flexibility to adjust weather, lighting, traffic, and geographic parameters. Second, the simulator experience can take an extensive period for a thorough evaluation of the system with zero damage and cost. Third, a real test drive may not always be feasible, for instance, during an auto show, a tech exhibition, or an academic conference, while the simulator is much easier to carry around and can be set up at different places. Overall, driving simulators, with unique advantages, can complement the goal of AV demonstration.

\subsection{Simulator for AV Education}
Driving simulators can also be used for driver’s education. We anticipate two types of users: those with a valid driver’s license and those without. During our study, many participants pointed out that just because someone holds a valid driver’s license does not mean they should be legally allowed to use an AV. Currently, due to limited knowledge and experience, drivers tend to blindly trust or mistrust AVs and often use them with overconfidence or little confidence. As a countermeasure, AV specified training and qualification would be recommended, and a driving simulator would make a solid platform for education delivery. For those without a driver’s license, a simulator can also be a good tool to practice driving skills, and the user can learn to drive a conventional vehicle while interacting with AV features.

\section{Conclusion}
We presented in this work a virtual reality simulator and an interactive driving experience to improve people’s understanding and trust of autonomous vehicles. To that effect, we adapted the open-source driving simulator CARLA and designed several scenarios. A study with 36 participants showed that our simulator successfully improved their attitude towards autonomous vehicles in terms of trust, perceived risk, perceived usefulness, perceived ease-of-use, and correspondingly behavioral intention. Several limitations exist. First, our small number of participants cannot represent all age, gender, and population groups. Future study will require a greater number of participants with varying backgrounds to achieve more comprehensive and objective analysis. Also, the simulation environment needs to be enhanced to support higher fidelity and user interactivity. Third, our customized AV stack may not be a faithful representation of a real vehicle and will require further development. Overall, We presented the rationale of using a simulator at driving schools, auto dealerships, and other places. The simulator and human study presented in this paper can act as an innovative, pioneering effort to promote safe autonomous vehicle education, training, and demonstration.

\bibliography{references.bib}

\end{document}